\documentclass[11pt,a4paper]{article}
\pdfoutput=1
\usepackage{jheppub}
\usepackage{amsfonts, amssymb, amsmath, amsthm}
\usepackage{mathrsfs}
\usepackage{graphicx}
\usepackage[utf8]{inputenc}
\usepackage[english]{babel}
\usepackage{slashed}
\usepackage{tikz-cd}

\usepackage{color}
\definecolor{dark-gray}{gray}{0.20}
\definecolor{gray}{gray}{0.30}
\definecolor{light-gray}{gray}{0.80}
\definecolor{dark-red}{rgb}{0.7,0,0}
\definecolor{dark-green}{rgb}{0.1,0.4,0}
\definecolor{dark-blue}{rgb}{0.3,0.3,0.7}
\definecolor{light-blue}{rgb}{0.8,0.8,1}
\definecolor{blue}{rgb}{0,0,1}
\definecolor{red}{rgb}{1,0,0}
\definecolor{green}{rgb}{0,1,0}

\usepackage{hyperref}
\hypersetup{
	colorlinks=true,
	linkcolor=dark-blue,
	citecolor=dark-red,
	urlcolor=dark-blue,
	linktoc=page
}

\def\cI{{\cal I}}

\def\cC{{\cal C}}

\def\sl{\mathfrak{sl}}
\def\i{{\rm i}}

\theoremstyle{definition}

\theoremstyle{identity}

\theoremstyle{remark}

\newcommand{\be}{\begin{equation}}
\newcommand{\ee}{\end{equation}}
\newcommand{\ba}{\begin{aligned}}
\newcommand{\ea}{\end{aligned}}
\newcommand{\bea}{\begin{eqnarray}}
\newcommand{\eea}{\end{eqnarray}}

\newcommand{\mathe}{\mathrm{e}}

\newcommand{\e}{\epsilon}
\newcommand{\lam}{\lambda}
\newcommand{\BV}{\mathbb{V}}
\newcommand{\BT}{\mathbb{T}}

\newcommand{\BC}{\mathbb{C}}
\newcommand{\BR}{\mathbb{R}}

\title{On M2-, M5-, and D3-branes from 12d anomalies:\\ an equivariant perspective on duality}

\author[a,b,c]{Kiril Hristov}
\affiliation[a]{Faculty of Physics, Sofia University, J. Bourchier Blvd. 5, 1164 Sofia, Bulgaria~\footnote{on leave}}
\affiliation[b]{INRNE, Bulgarian Academy of Sciences, Tsarigradsko Chaussee 72, 1784 Sofia, Bulgaria}
\affiliation[c]{School of Mathematics and Physics, University of Queensland,\\ St Lucia, Brisbane, Queensland 4072, Australia~\footnote{visiting}}

\author[d]{and Pedro Vicente Marto}
\affiliation[d]{Institute for Theoretical Physics, Utrecht University, 3584 CE Utrecht, The Netherlands}

\abstract{
\noindent We revisit the twelve-dimensional anomaly polynomial that has long been known to govern M5-brane worldvolume theories, together with its more recently proposed D3-brane counterpart. We show that these constructions naturally fit into the framework of equivariant integration, allowing the anomaly inflow formulas to be rewritten entirely in terms of equivariant volumes. In particular, we use the fact that the Bott--Cattaneo formula for the integral of powers of the global angular form on $S^{2k}$ emerges from the equivariant Euler-class integration of $\BC^k$, with two copies glued antipodally at the north and south poles of the sphere. Building on this perspective, we generalize the anomaly construction to include both electrically and magnetically charged branes, thereby incorporating M2-branes as well as a dual formulation of D3-branes. This provides a unified framework encompassing a number of classical results together with several recent developments, and offers a natural interpretation of the recently proposed M2/M5 duality, \cite{Hristov:2026tde}, in terms of an AGT-like correspondence. Remarkably, it also leads to a direct reinterpretation of constant-map contributions in topological string theory as manifestations of M-theory anomalies.
}
\date{\today}
\makeatletter
\gdef\@fpheader{}
\makeatother

\begin{document}
\maketitle

\section{Inroduction and summary of results}

A recurring theme in holography is that protected observables of brane worldvolume theories can be described in terms of geometric data of the ambient spacetime. Rather than requiring a direct analysis of the interacting worldvolume theory, quantities such as supersymmetric partition functions, indices, and on-shell actions can often be recast as topological or equivariant quantities and evaluated by localization. This perspective has appeared in a variety of seemingly distinct settings, suggesting a common geometric framework underlying many exact results. The purpose of this paper is to make this connection explicit for M2-, M5-, and D3-branes. In particular, we formulate several such constructions in terms of equivariant volumes and use this perspective to clarify the recently proposed relation between M2- and M5-brane partition functions, \cite{Hristov:2026tde}.

A classical realization of this viewpoint is provided by anomaly inflow, see \cite{Alvarez-Gaume:2022aak} for a recent review. The anomalies of a $p$-brane worldvolume theory are encoded in a $(p+3)$-form anomaly polynomial, whose inflow is generated by bulk Chern--Simons-like couplings~\cite{Duff:1995wd,Witten:1996md,Witten:1996hc,Freed:1998tg}. For M5-branes, this construction was developed by Harvey, Minasian and Moore~\cite{Harvey:1998bx}, and has since become an important tool for extracting exact 't~Hooft anomalies of supersymmetric field theories obtained from wrapped M5-branes~\cite{Bah:2018jrv,Bah:2019jts,Bah:2019rgq} and, more recently, wrapped D3-branes in type IIB~\cite{Bah:2020jas}.\footnote{Here and in the following, we use the term \emph{anomaly} in the sense of anomaly-inflow data: characteristic-class polynomials encoding the dependence of a brane worldvolume theory on background gauge and gravitational fields. This terminology does not imply an uncancelled anomaly in the underlying ten-dimensional string theory. In particular, while the local ten-dimensional anomalies of type IIB cancel, its D-branes possess non-trivial anomaly-inflow polynomials. The twelve-dimensional form proposed below should therefore be understood as an extension of the characteristic-class data governing the D3-brane, rather than as a ten-dimensional anomaly polynomial of type IIB itself.} Closely related structures also govern the Cardy limit of superconformal indices on compact Sasaki--Einstein backgrounds~\cite{Kim:2012ava,Assel:2015nca,Bobev:2015kza,Brunner:2016nyk,Nahmgoong:2019hko,Ohmori:2021dzb,Hosseini:2021fge,Cassani:2024tvk}. As we explain below, these constructions admit a natural interpretation in terms of equivariant integration, in which the anomaly polynomial becomes a higher-dimensional geometric object whose reductions give rise to several lower-dimensional observables.

A complementary development has emerged from the gravitational path integral itself. Beginning with the gravitational-block description of holographic on-shell actions and black-hole entropies~\cite{Hosseini:2019iad,BenettiGenolini:2019jdz,BenettiGenolini:2023kxp}, equivariant localization has been extended to a broad class of supergravity theories and spacetime dimensions~\cite{BenettiGenolini:2024lbj,BenettiGenolini:2024kyy,BenettiGenolini:2026qdm,Galli:2026lsl}, including formulations in terms of local patches and boundary contributions~\cite{Cassani:2024kjn,Colombo:2025ihp,Colombo:2025yqy}. From a complementary geometric perspective, equivariant localization has also been developed for non-compact toric Calabi--Yau cones~\cite{Martelli:2023oqk,Colombo:2023fhu}, see also \cite{Nekrasov:2021ked,Cassia:2022lfj}. Together, these developments point toward a common geometric language relating anomaly inflow, holographic localization, and toric geometry.

The same geometric perspective emerges on the M2-brane side from equivariant topological strings. It was observed that higher-derivative $\mathcal N=2$ supergravity reproduces the perturbative finite-$N$ $S^3_b$ partition functions of ABJM-like theories~\cite{Bobev:2021oku,Hristov:2021qsw,Hristov:2022lcw}. A formulation in terms of equivariant volumes was developed in~\cite{Cassia:2025aus} and applied to M2-brane partition functions in~\cite{Cassia:2025jkr}, relating to the extensive literature on supersymmetric localization in three dimensions, ~\cite{Kapustin:2009kz,Drukker:2010nc,Hama:2011ea,Marino:2011eh,Hatsuda:2013oxa,Nosaka:2015iiw,Hatsuda:2016uqa,Chester:2021gdw,Geukens:2024zmt,Kubo:2024qhq,Bobev:2025ltz,Hristov:2026zjh,Hong:2026zul} and \cite{Kim:2009wb,Kapustin:2011jm,Beem:2012mb,Benini:2015noa,Benini:2015eyy,Hosseini:2016tor,Hosseini:2016ume,Hosseini:2022vho,Bobev:2022eus,Bobev:2024mqw,Inglese:2023wky,Colombo:2024mts}. Motivated by these developments, Ref.~\cite{Hristov:2026tde} proposed that the anomaly polynomial and the topological-string free energy admit a common equivariant description, providing further evidence for the M2/M5 relation conjectured by Chen, Dorey, Moriyama, Mouland and Sanli~\cite{Chen:2026fpe},~\footnote{See also \cite{Batrachenko:2002pu} for an older proposal with a similar flavor.} itself inspired by earlier studies of giant graviton expansions~\cite{Gaiotto:2021xce,Imamura:2021ytr,Hayashi:2024aaf}. In this work, we develop this perspective further by showing that the underlying anomaly constructions can themselves be formulated directly in terms of equivariant volumes, thereby placing these developments within a unified geometric framework.

Concretely, the central object of this paper is the anomaly 12-form
$\cI_{12}$, integrated over a twelve-dimensional extension space $Y_{12}$ satisfying
$\partial Y_{12}=M_{11}$ (or $\partial^2Y_{12}=M_{10}$ in type IIB). We denote the resulting zero-form by
\be
\label{eq:I0intro}
\cI_0 :=
\pi_{Y*}\!\left(\cI_{12}\right)\ ,
\ee
where $\pi_Y:Y_{12}\rightarrow\mathrm{pt}$ is the projection to a point. Throughout this paper,
the pushforward is understood \emph{equivariantly}, namely through the localization formula
\be
\pi_{Y*}(\alpha)
=
\int_{Y_{12}}\alpha^{\BT} =
\sum_{p\in Y}
\frac{\alpha|_p}
{e^{\BT}(N_p)}\ , 
\ee
where the sum runs over the isolated fixed points,
$e^{\BT}(N_p)$
denotes the equivariant Euler class of the normal bundle to the fixed point, and $\alpha$ is an arbitrary characteristic form.

We interpret $\cI_0$ as the leading contribution to the ($\log$ of the) protected partition function in the
regime of small equivariant parameters, corresponding for example to the Cardy limit of the
M5-brane index or to the perturbative sector of the M2-brane free energy.~\footnote{Due to varying conventions, we are not going to attempt giving a precise normalization here between $\cI_0$ and $\log Z$ for any of the underlying partition functions.}

Suppose now that the extension space factorizes as
\be
Y_{12}\cong Z_d\times X_{12-d}\ ,
\ee
and let
\be
\pi_Z:Y_{12}\rightarrow X_{12-d}\ ,
\qquad
\pi_X:Y_{12}\rightarrow Z_d\ ,
\ee
denote the natural projections. These define intermediate equivariant classes
\be
\cI_{12-d}
=
\pi_{Z*}\!\left(\cI_{12}\right)\ ,
\qquad
\cI_d
=
\pi_{X*}\!\left(\cI_{12}\right)\ ,
\ee
while the full equivariant pushforward factorizes as
\be
\label{eq:pushforwardfactorization}
\pi_{Y*}
=
(\pi_{X\rightarrow\mathrm{pt}})_*\circ\pi_{Z*}
=
(\pi_{Z\rightarrow\mathrm{pt}})_*\circ\pi_{X*}\ ,
\ee
whenever such a product decomposition exists.

Although mathematically elementary, this factorization acquires a natural physical
interpretation for brane systems. The intermediate forms obtained by partial pushforward naturally encode characteristic data of lower-dimensional effective descriptions, related to the consistent truncations of supergravity and string theory, see e.g.\ \cite{Cvetic:2000dm,Lee:2014mla,Cassani:2019vcl}. The anomaly polynomial is built from the field strength
carrying the corresponding brane charge, and the associated flux distinguishes electric and
magnetic descriptions. For magnetically charged branes, the flux is measured through a sphere
linking the brane, so the first pushforward is naturally performed over the transverse
directions. For electrically charged branes, the field strength instead fills the worldvolume,
making it natural to integrate first over the parallel directions. Throughout this paper we
adopt the convention of performing the first equivariant pushforward along the directions
threaded by the corresponding form field.

Schematically, the two realizations of the equivariant anomaly polynomial are summarized by
\begin{equation}
\label{diagram:1}
\begin{tikzcd}[column sep=4.5em,row sep=3em]
& \cI_{12}
  \arrow[dl,"Z"']
  \arrow[dr,"Z"] & \\
\cI^{\rm el}_{12-d}(\mu_{\rm el};\omega)
  \arrow[d,"X"']
&&
\cI^{\rm mag}_{12-d}(N_{\rm mag};\omega)
  \arrow[d,"X"] \\
\cI^{\rm el}_{0}(\mu_{\rm el};\omega,\epsilon)
&&
\cI^{\rm mag}_{0}(N_{\rm mag};\omega,\epsilon)
\end{tikzcd}
\end{equation}
where $\omega$ and $\epsilon$ denote the equivariant parameters associated with the
factors $Z_d$ and $X_{12-d}$, respectively. Importantly, supersymmetry dictates that the full 12d geometry satisfies an equivariant Calabi-Yau condition for the first Chern class,~\footnote{To match some common conventions in the literature, we have implicitly introduced an additional minus sign for the equivariant parameters of $Z$.}
\be
\label{eq:susyintro}
    c_1^\BT (Y_{12}) = \sum_i \e_i - \sum_\alpha \omega_\alpha = 0\ .
\ee
The two resulting zero-forms are generally interpreted as partition function of two different brane systems that in addition belong to different thermodynamic ensembles, see \cite{Kurlyand:2022vzv,Beccaria:2023hhi,Gautason:2025plx,Bobev:2026gir}. The
left branch describes an electric realization, in which one fixes the chemical potential
$\mu_{\rm el}$ conjugate to the brane charge, whereas the right branch gives the magnetic
description at fixed charge $N_{\rm mag}$. We shall refer to the correspondence between the
two lower corners of Diagram~\eqref{diagram:1} as an \emph{equivariant
electric/magnetic duality}. It is important to emphasize that at this stage this is not a duality between
the underlying microscopic theories (even though it is natural to hypothesize that it holds at exact quantum level), but between two complementary polarizations of the same
equivariant anomaly computation. The emerging picture is clearly reminiscent of the AGT correspondence and its generalizations, \cite{Alday:2009aq,LeFloch:2020uop}.

The M2/M5 system provides a clean realization of this picture. The two descriptions are
obtained from the same equivariant twelve-dimensional geometry,
\be
Y_{12}=Z_4\times X_8\ ,
\ee
but with different interpretations of the two factors:
\begin{equation}
\begin{aligned}
\text{M2 (electric):}\qquad
&Y_{12}=\BC^2\times X_8^{(\perp)}\ ,\\
\text{M5 (magnetic):}\qquad
&Y_{12}=X_8^{(\parallel)}\times S^4\ .
\end{aligned}
\end{equation}
The compact sphere $S^4$ and its local equivariant model $\BC^2$ are related through the
Bott--Cattaneo formula, so that both constructions evaluate to the same equivariant
characteristic classes, but in different polarizations of the geometry. Consequently,
although the microscopic M2- and M5-brane theories are distinct, the protected quantities
computed from the anomaly polynomial satisfy the correspondence
\begin{equation}
\label{eq:mainduality}
\cI_0^{\rm M2}(\mu_{\rm M2};\omega,\epsilon)
\quad\longleftrightarrow\quad
\cI_0^{\rm M5}(N_{\rm M5};\omega,\epsilon)\ ,
\end{equation}
which is precisely the relation established in \cite{Hristov:2026tde} between the
perturbative M2-brane partition function and the Cardy limit of the M5-brane index. The
present work identifies this relation as the first example of a more general equivariant
electric/magnetic duality, expected whenever two electromagnetically dual brane systems can
be described by different polarizations of the same equivariant anomaly polynomial.

\medskip
\noindent\textbf{Aim of this paper.}
The present work has two main goals. The first is to reformulate the reduction
$\cI_{12}\longrightarrow \cI_0$ directly at the level of the M-theory anomaly polynomial,
\be
\label{eq:Manom}
\cI_{12}^{\rm M}
=
\frac16\,E_4^3
+
E_4\wedge I_8\ ,
\ee
with $I_8$ the 8-derivative term that completes the effective action, \cite{Witten:1996hc,Freed:1998tg,Harvey:1998bx}.
We show that the compact equivariant pushforward over a sphere reproduces the classical
Bott--Cattaneo formula for the global angular form~\cite{1997dg.ga}, see \eqref{eq:BCformula}, and explain how this identity follows naturally from equivariant localization.
The sphere is viewed as the gluing of two equivariant copies of $\BC^k$ centered at the
north and south poles, carrying opposite equivariant parameters. The Bott--Cattaneo
formula is then recovered by summing the two fixed-point contributions, in complete
agreement with the localization picture reviewed in Section~3.1.2 of
\cite{Hosseini:2020vgl}. This provides an equivariant interpretation of the
Harvey--Minasian--Moore anomaly inflow construction for M5-branes and, more importantly,
extends immediately to non-compact geometries, e.g.\ where only a single equivariant fixed point
contributes.

The resulting M2-brane computation is one of the central observations of this work.
Performing the equivariant pushforward over the (Euclidean) AdS$_4$ directions produces an
eight-form anomaly polynomial $\cI_8^{\rm M2}$ on the transverse Calabi--Yau fourfold.
We show that this polynomial agrees precisely with the equivariant characteristic-class
expression governing the perturbative constant-map sector of the topological-string
partition function derived in
\cite{Cassia:2025aus,Cassia:2025jkr}. In this way, a quantity previously obtained from
equivariant topological strings is recovered directly from the classical M-theory anomaly
polynomial. This agreement strongly suggests that the perturbative topological-string
partition function should be viewed as a quantum refinement of the underlying classical anomaly integral.\\

Our second goal is to show that the same equivariant framework extends naturally to
type IIB string theory. Motivated by the anomaly-inflow construction of
Bah, Bonetti, Minasian and Weck~\cite{Bah:2020jas}, we argue that we should consider the
twelve-dimensional form
\be
\label{eq:Banom}
\cI_{12}^{\rm IIB}
= \frac12\,
E_6\wedge E_6\ ,
\ee
relevant specifically for D3-branes. Because the type IIB five-form field strength is self-dual, D3-branes admit both magnetic
($N$-ensemble) and electric ($\mu$-ensemble) descriptions. The corresponding equivariant
pushforwards are summarized schematically by
\begin{equation}
\label{diagram:2}
\begin{tikzcd}[column sep=4.5em,row sep=3em]
& \cI_{12}^{\rm IIB}
  \arrow[dl,"X"]
  \arrow[dr,"Z"] & \\
\cI^{\rm D3}_6(\mu;\epsilon)
  \arrow[d,"Z"]
&&
\cI^{\rm D3}_6(N;\omega)
  \arrow[d,"X"] \\
\cI^{\rm D3}_0(\mu;\omega,\epsilon)
  \arrow[rr,"\text{Legendre}",bend left=15]
&&
\cI^{\rm D3}_0(N;\omega,\epsilon)
  \arrow[ll,"\text{inverse Legendre}",bend left=15]
\end{tikzcd}
\end{equation}
where
\be
Y_{12}=X_6^{(\parallel)}\times Z_6^{(\perp)}\ ,
\ee
is decomposed into worldvolume and transverse Calabi--Yau threefolds.
The two descriptions lead to~\footnote{In section \ref{sec:4.3} we have included a brief discussion on how the main results below can be unpacked to match the usual cubic polynomials that appear more often in literature.}
\be
\cI_0^{\rm D3}(\mu)
=-
\frac{\mu^2}{2}\,
\frac{C_Z(\omega)}{C_X(\epsilon)}\ ,
\qquad
\cI_0^{\rm D3}(N)
=
\frac{N^2}{2}\,
\frac{C_X(\epsilon)}{C_Z(\omega)}\ ,
\ee
which are related by a Legendre transform exchanging the two ensembles. The quantities $C_X, C_Z$ featuring prominently above are simply the equivariant pushforwards of the identity class on the respective manifolds, see \eqref{eq:mesonicvol}, which can be considered as the most basic building blocks of any explicit calculation. These expressions reproduce the large-$N$
partition functions of $\mathcal N=4$ SYM and any other theory of D3-branes on toric Calabi-Yau cones. In the $N$-ensemble they recover the well-known
relation between $a$-maximization and Sasakian volume minimization, see \cite{Martelli:2005tp,Butti:2005vn,Butti:2005ps,Martelli:2006yb,Amariti:2011uw,Couzens:2018wnk,Gauntlett:2018dpc,Hosseini:2019use,Hosseini:2019ddy,Gauntlett:2019roi,Gauntlett:2019pqg,Boido:2022mbe}, while in the $\mu$-ensemble they coincide
with the extremization principle proposed in
\cite{Martelli:2023oqk,Colombo:2023fhu}.

Finally, the D3-brane construction provides a useful consistency check of the general
equivariant electric/magnetic correspondence proposed in this paper. Unlike the M2/M5 system,
where the electric and magnetic descriptions relate distinct brane theories, the two
descriptions now arise from the same self-dual D3-brane and are therefore related simply by a
Legendre transform. However, we still observe a leftover electric/magnetic duality when the worldvolume and transverse equivariant geometries are identified,
\be
X=Z\ ,\qquad \epsilon_i=\omega_i\ .
\ee
In this case diagrams~\eqref{diagram:1} and~\eqref{diagram:2} coincide, and one
finds a direct relation
\be\label{eq:D3duality}
\cI_0^{\rm D3}(N;\e,\epsilon)
=
\cI_0^{\rm D3}(\mu = \pm \i\,N;\e,\epsilon)\ ,
\ee
which may be viewed as the D3 counterpart of the M2/M5 correspondence. It would be interesting to extend this correspondence to the non-conformal branes in type IIB, which would be a natural continuation of the present work.

\medskip
\noindent\textbf{Outline.} The rest of the paper is organized as follows. In
Section~\ref{sec:review} we set up conventions for the equivariant volume of toric Calabi--Yau
cones and rederive the Bott--Cattaneo formula equivariantly, following \cite{Hosseini:2020vgl}. We emphasize the similarities and differences between integration on even and odd spheres and their distinct relation to equivariant volume calculations, relevant respectively for the M5 and D3 branes. In Section~\ref{sec:Mtheory} we
carry out the M5- and M2-brane anomaly-polynomial integrals from the M-theory polynomial, reviewing and completing the derivation of~\cite{Hristov:2026tde}. In
Section~\ref{sec:IIB} we perform the analogous calculation for D3-branes in both the electric and magnetic frames. We conclude in Section~\ref{sec:discussion}
with a discussion of open problems. An appendix collects our conventions and explicit examples for equivariant characteristic
classes, following~\cite{Cassia:2025aus,Hristov:2026tde}.

\medskip
\noindent\textbf{Note added.} While this paper was being finalized, several interesting references appeared: \cite{Cassani:2026teb,Couzens:2026xmi,BenettiGenolini:2026cdw,BenettiGenolini:2026cyc}. They all touch on closely related subjects and are in agreement with the present results where they overlap, but their starting points are different, and therefore the present work can be considered complementary. We have included further comments in the concluding section.

\section{Equivariant volumes and the Bott--Cattaneo formula}
\label{sec:review}

We collect here some equivariant geometry facts used throughout the paper, following
\cite{Cassia:2025aus,Martelli:2023oqk}, and use them to give a self-contained equivariant
derivation of the classical Bott--Cattaneo integration formula
\cite{1997dg.ga} for global angular forms on even-dimensional sphere bundles. We also comment on the odd-dimensional sphere case, which is conceptually different.

\subsection{Equivariant volumes of $\BC^n$ and $S^2$}

For a toric K\"ahler quotient of complex dimension $d=n-r$,
\be
X=\BC^n/\!\!/U(1)^r\ ,
\ee
 the \emph{equivariant volume} is,~\footnote{Note that we use the terminology equivariant volume somewhat incorrectly, referring more generally to a generating function.} in the conventions of~\cite{Cassia:2025aus},
\be
\label{eq:VXdef}
\BV_X(\lambda,\epsilon)
:= (\pi_{X \to \text{pt}})_* (1) = 
\int_X e^{\omega_\lambda^{\BT}}
=
\oint_{\rm JK}
\prod_{a=1}^r
\frac{d\phi_a}{2\pi i}\,
\frac{e^{x_i\lambda^i}}
{\prod_{i=1}^n x_i}\ ,
\qquad
x_i=\epsilon_i+\phi_aQ_i^a\ ,
\ee
where $\omega_\lambda^\mathbb{T}$ is the equivariant completion of the symplectic form $\omega_\lambda=\sum_a\lambda_a\phi^a$, see \cite{Martelli:2023oqk},  and $Q_i^a$ are the GLSM charges defining the quotient. The contour is fixed by the
Jeffrey--Kirwan (JK) prescription, or equivalently by equivariant localization onto the fixed points. See \cite{Cassia:2025aus} for more details. The equivariant volume~\eqref{eq:VXdef} is a generating function for equivariant intersection
numbers, and in turn characteristic classes. In particular, differentiation with respect to the K\"ahler parameters inserts powers of
the equivariant Chern roots $x_i$, see the Appendix for more details. In the simplest case when there are no insertions, evaluating the equivariant volume at vanishing K\"ahler parameters corresponds to the equivariant pushforward of the identity class (which was dubbed \emph{mesonic} volume in \cite{Cassia:2025aus}):
\be
\label{eq:mesonicvol}
    C_X (\e) := (\pi_{X \to \text{pt}})_* (1) = \BV (\lam = 0, \e)\ .
\ee

The simplest example is $X=\BC^n$, for which there is no gauging ($r=0$) and therefore
\be
\label{eq:Cvol}
\BV_{\BC^n}(\lambda,\epsilon)
=
\frac{e^{\epsilon_i\lambda^i}}
{\prod_{i=1}^n\epsilon_i}\ .
\ee
As the simplest compact example, consider $S^2\simeq \mathbb{CP}^1$,
realized as the $U(1)$ quotient of $\BC^2$ with GLSM charges $Q=(1,1)$.
Evaluating~\eqref{eq:VXdef} by residues at the two fixed points
($x_1=0$ and $x_2=0$) gives
\be
\label{eq:VS2}
\BV_{S^2}(\lambda,\omega)
=
\frac{e^{\omega \lambda^2}-e^{-\omega\lambda^1}}
{\omega}\ ,
\qquad
\omega:=\epsilon_2-\epsilon_1\ .
\ee
The two terms are precisely the contributions of the north and south poles, where the tangent
space is locally identified with $\BC$ carrying equivariant weights $\pm\omega$, respectively.
Equation~\eqref{eq:VS2} therefore makes manifest that the equivariant volume of $S^2$ is obtained
by gluing two local copies of $\BC$ with opposite equivariant weights, while keeping the (redundant) K\"ahler
parameters independent. Note that the description of $\mathbb{V}_{S^2}$ in terms of $\mathbb{V}_\mathbb{C}$ implicitly involved a reparametrization of the equivariant parameters $\epsilon_i$ into the weights of the toric action at the north/ south pole fixed points, $\omega=\pm(\epsilon_2-\epsilon_1)$.

\subsection{Bott--Cattaneo and even-dimensional spheres}
\label{subsec:even}

The global angular form $e_{2k}$ is an $SO(2k+1)$-invariant differential
form on the unit sphere bundle $S(E)$ of an oriented rank-$(2k+1)$ vector bundle $E$,
\be
\begin{array}{ccc}
 \mathbb{R}^{2k+1} & \longrightarrow & E \\
 && \downarrow{\pi} \\
 && B
\end{array}
\qquad\qquad
\begin{array}{ccc}
 S^{2k} & \longrightarrow & S(E) \\
 && \downarrow{\pi} \\
 && B
\end{array}
\ee
normalized such that~\footnote{Here and in the formula below, we use the integral sign instead of the pushforward notation in order to emphasize that this corresponds to an ordinary, non-equivariant, integration.}
\begin{equation}
\label{eq:normalizationangular}
    \int_{S^{2k}} e_{2k}=2\ .
\end{equation}
Since $E$ has odd rank, its Euler class vanishes identically and the global
angular form is closed,
\begin{equation}
    {\rm d}e_{2k}=0\ .
\end{equation}
The Bott--Cattaneo formula gives the fiber integrals of arbitrary powers of the
global angular form \cite{1997dg.ga,Harvey:1998bx,Freed:1998tg},
\begin{equation}
\label{eq:BCformula}
    \int_{S^{2k}}\left(e_{2k}\right)^{\,2s+1}
    =2\,p_k(E)^s\ ,
    \qquad
    \int_{S^{2k}}\left(e_{2k}\right)^{\,2s}=0\ ,
\end{equation}
where $p_k(E)$ denotes the $k$-th Pontryagin class of $E$.

The global angular form provides a differential-form representative of the
transgression of the Thom class of the vector bundle, which will provide a
useful physical interpretation in the next section. Choosing a Thom form
$U(E)$ with compact support in the fiber directions, one may write away from
the zero section
\begin{equation}
\label{eq:Thomform}
    U(E)={\rm d}\!\left(\rho\,e_{2k}\right)\ ,
\end{equation}
where $e_{2k}$ is implicitly pulled back to $E\setminus B$ and
$\rho(r)$ is a smooth radial cutoff function. The Euler class of $E$ is then
obtained by restricting the Thom class to the zero section,
\begin{equation}
    e(E)=s^*U(E)\ .
\end{equation}
This relation provides the link between the global angular form and the
Euler class appearing in the original version of the Bott--Cattaneo formula, \cite{1997dg.ga}.

We now want to rederive the Bott-Cattaneo formula using the equivariant volume discussed previously. Following \cite{Hosseini:2020vgl}, we argue that this amounts to \emph{equivariant} integration of the Euler class of the sphere itself, $e(TS^{2k})$. Concretely, we can simply take the case $k=1$, such that for $S^2$ we can imediately use \eqref{eq:VS2},
\be
\pi_*
\Big((e(TS^2))^m\Big)
=
(\partial_{\lambda^1}+\partial_{\lambda^2})^m\,
\BV_{S^2}(\lambda,\omega)
\Big|_{\lambda=0}
=
\begin{cases}
2\,\omega^{2s}\ ,
&
m=2s+1\ ,
\\[2mm]
0\ ,
&
m=2s\ .
\end{cases}
\label{eq:S2BC}
\ee
Equation~\eqref{eq:S2BC} may be viewed as the simplest equivariant incarnation of the
Bott--Cattaneo formula. In the non-equivariant limit $\omega\rightarrow0$ it reduces to the familiar
identity $ \int_{S^2}e(TS^2)=\chi(S^2)=2$ for $m=1$, while all higher powers vanish because their ordinary differential-form degree exceeds
the dimension of the manifold. At finite equivariant parameters, however, the higher odd powers
remain non-trivial and evaluate to the polynomial $2\omega^{2s}$, reflecting the fact that the
equivariant Euler class is an element of equivariant cohomology rather than ordinary de~Rham
cohomology~\cite{Hosseini:2020vgl}.~\footnote{Although $(e(TS^2))^{m,\BT}$ has total degree larger than the real dimension of $S^2$
for $m>1$, equivariant integration remains well-defined because the equivariant Euler class is a
polynomial in the equivariant parameters with coefficients in ordinary differential forms. The
localization theorem computes the corresponding pushforward exactly from the contributions of the
two fixed points, yielding~\eqref{eq:S2BC}.} Intiuitively, this is analogous to the result encapsulated in the Bott--Cattaneo formula \eqref{eq:BCformula}, which produces a non-trivial result even when the rank of a given power of the global angular form exceeds the dimension of the sphere $S^{2k}$ over which it is integrated. This is only possible if the volume form on $S^{2k}$ is gauged over the base manifold and there is a non-trivial $SO(2k+1)$ connection. This ensures that $e_{2k}^{2s+1}$ has legs on the base manifold and its integral over the sphere is non-vanishing. The relation between both computations is explained more clearly below.

The same gluing construction extends naturally to higher even-dimensional spheres. Although
$S^{2k}$ with $k>1$ is not itself a toric K\"ahler manifold, and hence the integral
representation~\eqref{eq:VXdef} does not apply globally, each of the two neighborhoods of the
north and south poles is equivariantly modeled on $\BC^k$. As in the $S^2$ example, the full
sphere is therefore reconstructed by gluing these two local patches with opposite equivariant
weights, following the argument outlined in~\cite[Section~3.1.2]{Hosseini:2020vgl}. In particular, $S^{2k}$ with its standard $U(1)^k$ action admits two fixed points, the north and south poles, where we regard $TS^{2k}|_p=\oplus_jL_j$ as a sum of complex line bundles.

Let us then define
\be
\mathcal D_k
:=
\sum_{i_1<\cdots<i_k}
\partial_{\lambda^{i_1}}\cdots
\partial_{\lambda^{i_k}}
=
\prod_{i=1}^{k}\partial_{\lambda^i},
\ee
which inserts the equivariant top Chern class
$c_k^{\BT}=e^{\BT}(T\BC^k)$.
Acting with $(\mathcal D_k)^m$ on the equivariant volume of $\BC^k$
and summing the two pole contributions with the relative sign
$(-1)^k$ induced by the orientation reversal of the fiber allows us to define
\bea
\label{eq:SviaC}
\pi_*
\Big((e(TS^{2k}))^m\Big)
& :=
\left[
\mathcal (\mathcal{D}_k)^m
\BV_{\BC^k}(\lambda;\omega_1,...,\omega_k)
\right]_{\lambda=0}
+
(-1)^k
\left[
\mathcal (\mathcal{D}_k)^m
\BV_{\BC^k}(\lambda;-\omega_1,...,-\omega_k)
\right]_{\lambda=0}
\notag\\
&=
\begin{cases}
2(\omega_1\cdots\omega_k)^{2s},
&
m=2s+1,
\\[2mm]
0,
&
m=2s,
\end{cases}
\label{eq:BCgeneral}
\eea
where we have written the equivariant parameters as $\omega_i$ to emphasize the analogy with the
$S^2$ example.\footnote{As in the $S^2$ example, we have the identity $\pi_*(e(TS^{2k}))=2$, which is the equivariant generalization of the normalization defined in \eqref{eq:normalizationangular} for the integral of the global angular form over the fiber $S^{2k}$.} 

The relation between~\eqref{eq:BCgeneral} and the classical Bott--Cattaneo formula follows from
Chern--Weil theory. We need to assume that that
structure group of $E$ is reduced to the maximal torus
$U(1)^k\subset SO(2k+1)$, such that after complexification
$E$ splits locally into complex line bundles
$L_1\oplus\cdots\oplus L_k$,
whose Chern roots are represented by the equivariant parameters
$\omega_i$.
Under the Chern--Weil homomorphism,
\be
\label{eq:CW1}
\omega_i
\longmapsto
\frac{F_i}{2\pi}\ ,
\ee
with $F_i$ the curvature of $L_i$, symmetric polynomials in the $\omega_i$ are mapped to the
corresponding characteristic classes of $E$. In particular,
\be
\label{eq:CW2}
(\omega_1\cdots\omega_k)^2
\longmapsto
p_k(E)\ .
\ee
Applying the Chern--Weil map to~\eqref{eq:BCgeneral} therefore immediately yields \eqref{eq:BCformula},
which is precisely the Bott--Cattaneo formula, \cite{1997dg.ga}, later used by Harvey, Minasian and Moore in the derivation of the M5-brane
anomaly polynomial~\cite{Harvey:1998bx}.

The mapping \eqref{eq:CW2} also explains how the result of the equivariant integration \eqref{eq:SviaC}, dependent only on equivariant parameters, is translated into a genuine differential form resulting from the Bott-Cattaneo formula. To understand this, one can view the parameters $\omega_i$ as the equivariant completion of each field strength $F_i$ evaluated at a fixed point.\footnote{Restricting to the maximal torus, we can write $F_i=\begin{pmatrix} 0&x_i\\ -x_i & 0\end{pmatrix}$, with $x_i$ a 2-form, as $x_i\mapsto x_i+\omega_i$; this reduces to $\omega_i$ when evaluated at a fixed point.}

We stress that~\eqref{eq:BCgeneral} is fundamentally an identity in equivariant cohomology and
does not rely on the compactness of the fiber. Once expressed in terms of equivariant
pushforwards, the same computation extends immediately to non-compact geometries, where only the
relevant fixed-point contributions remain. This observation will be the starting point for our
equivariant treatment of M2-branes in the following sections, and has already been extensively used in \cite{Hosseini:2020vgl}.

\subsection{Odd-dimensional spheres}
\label{subsec:odd}

We now follow the analogous construction for even-dimensional spheres, emphasizing the important differences. This is of notable use when discussing the D3-brane picture. The global angular form $e_{2k-1}$ is an $SO(2k)$-invariant differential
form on the unit sphere bundle of an oriented rank-$2k$ vector bundle,
\be
\begin{array}{ccc}
 \mathbb{R}^{2k} & \longrightarrow & E \\
 && \downarrow{\pi} \\
 && B
\end{array}
\qquad\qquad
\begin{array}{ccc}
 S^{2k-1} & \longrightarrow & S(E) \\
 && \downarrow{\pi} \\
 && B
\end{array}
\ee
with normalization
\begin{equation}
    \int_{S^{2k-1} }e_{2k-1}=2\ .
\end{equation}
In contrast to the previous case, the Euler class of an even-rank bundle need
not vanish, and the angular form is not closed. Instead, it obeys
\begin{equation}
\label{eq:oddnormalization}
    {\rm d}e_{2k-1}=-\pi^*e(E)\ ,
\end{equation}
up to convention-dependent sign. Therefore there is no direct analogue of the
Bott--Cattaneo identities for powers of a closed angular form. Rather, the odd
sphere bundle is naturally viewed as the boundary of the disk bundle,
\begin{equation}
    S(E)=\partial D(E),
\end{equation}
and the angular form is again obtained as the transgression of the Thom form,
\begin{equation}
    U(E)=\text{d}\!\left(\rho\,e_{2k-1}\right),
\end{equation}
where $e_{2k-1}$ is implicitly pulled back to $E\setminus B$ and
$\rho(r)$ is a smooth radial cutoff function. The Euler class is again recovered
from the Thom class through
\begin{equation}
    e(E)=s^*U(E)\ .
\end{equation}

This perspective naturally replaces fiber integrals over the odd-dimensional
sphere by integrals of characteristic classes over the corresponding disk
bundle, which is of course in one higher dimension. In the equivariant setting, we are then lead to perform equivariant pushforward over the cone~\footnote{Since the boundary does not contribute to equivariant integration, there is no distinction between the disk and the cone for the present purposes.}
\be
C(S^{2k-1})\simeq \BC^{k},
\ee
where the equivariant Euler class of the tangent bundle
$T\BC^k$ enters naturally through localization. Fully explicitly, we evaluate
\be
\label{eq:CEulerclass}
    2\, \pi_* \Big((e(T\BC^{k}))^m\Big) = 2\, (\omega_1\cdots\omega_k)^{m-1}\ ,
\ee
where the prefactor was included to keep the normalization in \eqref{eq:oddnormalization}.
This provides the
equivariant analog of the Bott--Cattaneo construction for
even-dimensional sphere bundles, implicitly using once again the Chern-Weil map, see \eqref{eq:CW1}-\eqref{eq:CW2}. In particular, note that in this case the result is non-vanishing for even values of the power $m$, in contrast with \eqref{eq:BCformula}. This will be crucial when discussing the anomaly polynomial of type IIB theory.

\subsection*{Summary}
The relation between the two main constructions so far can be summarized as follows:
\begin{equation}
\begin{array}{c|c}
\text{Even-dim.} & \text{Odd-dim.} \\[2mm]
\hline
S^{2k}\subset \mathbb{R}^{2k+1} &
 C(S^{2k-1})\simeq\BC^{k} \\[2mm] \displaystyle
 \sum_{\sigma = \text{NP, SP}} \pi_* \left( (e(T\BC^k))^m \right)_\sigma \mapsto \int_{S^{2k}}\!\left(e_{2k}\right)^m &
\displaystyle 2\, \pi_* \left(e (T\BC^k))^m\right) \mapsto \int_{S^{2k-1}}\!\left(e_{2k-1}\right)^m 
\end{array}
\end{equation}

\section{M-theory: M5- and M2-brane anomalies}
\label{sec:Mtheory}

The anomaly inflow for spacetime-filling M2- and M5-branes in eleven-dimensional supergravity is
encoded in the twelve-form~\cite{Witten:1996hc,Freed:1998tg,Harvey:1998bx}
\be
\label{eq:I12M}
\cI_{12}^{\rm M}
=
\frac16\,E_4\wedge E_4\wedge E_4
+
E_4\wedge I_8,
\ee
where
\be
I_8
=
\frac1{192}
\bigl(p_1(TM)^2-4p_2(TM)\bigr)\ ,
\ee
is the familiar eight-form constructed from the Pontryagin classes of the eleven-dimensional
spacetime tangent bundle, $M = M_{11}$. In the original M5-brane construction, to which we turn now, the four-form $E_4$ encodes the topology of the bundle transverse to the brane and provides the differential-form representative entering the anomaly-inflow
construction.

\subsection{M5-branes}
\label{sec:M5}

Consider a stack of $N_\text{M5}$ coincident M5-branes with six-dimensional worldvolume $B_6$. The local
geometry near the branes is described by the rank-five normal bundle $\mathbb{R}^5 \hookrightarrow M_{11} \longrightarrow B_6$ with the M5-branes sitting at the origin of each fiber. We are thus clearly in the setting described in subsection \ref{subsec:even}.

The M5-brane are magnetically charged under the M-theory three-form potential and therefore
source $N_\text{M5}$ units of $G_4$ flux through the linking four-sphere,
\be
\frac1{2\pi}\int_{S^4}G_4=N_\text{M5}\ .
\ee
The main observation of \cite{Harvey:1998bx,Freed:1998tg} is that the above magnetic charge leads to a singular flux, with the singularity removed by excising a tubular neighborhood of the stack of branes, such that $M_{11}$ acquires a boundary, $\partial M_{11} = M_{10}$.

If $r$ denotes a radial function away from the M5-branes, the four-form flux then takes the form
\be
    \frac{G_4}{2\pi} = - \rho\, E_4 + \dots\ ,
\ee
where $\rho$ is a bump function interpolating between a constant in the tubular neighborhood and zero in the far-away region. The ellipses stand for possible subleading terms in the small $r$ limit, and $E_4$ is a closed and globally defined four-form on $M_{10}$. 

In the anomaly inflow construction this flux is represented geometrically by the four-form~\footnote{The present conventions might differ slightly from some of the previous literature, due to the normalization in \eqref{eq:normalizationangular}. For a clearer comparison, note that $\int_{S^4} E_4 = N_\text{M5}$.}
\be
E_4=\frac{N_\text{M5}}2\, e_4 ,
\ee
where $e_4$ is the global angular form on the sphere bundle $S^4\hookrightarrow M_{10}\longrightarrow B_6$. At this stage we can also notice the immediate relation between the four-form flux above and the Thom form, c.f.\ \eqref{eq:Thomform}.

The main observation here is that the same construction admits a natural equivariant
description, which was essentially already spelled out in subsection \ref{subsec:even}. The equivariant localization of the $S^4$ fiber is controlled by its two fixed
points, whose local neighborhoods are modeled by $\BC^2$, as in \eqref{eq:SviaC}. To fully relate with the notation we used in the Introduction, we should identify
\be
   \label{eq:M5split} Y_{12}^\text{M5}=X_8^{(\parallel)}\times Z_4^{(\perp)}
\simeq
X_8 \times S^4\ .
\ee
where $X_8$ is an auxiliary space formally extending the spacetime $B_6$, $\partial X_8 = B_6$. We can thus use the ansatz 
\be
    E_4 = \frac{N_\text{M5}}2\, e(TS^{4})\ ,
\ee
following the logic of the previous section. We can then use the intermediate results
\be
    \pi_{Z*} (E_4) = N_\text{M5}\ , \qquad \pi_{Z*} ((E_4)^3) = \frac{(N_\text{M5})^3}4\, (\omega_1 \omega_2)^2\ ,
\ee
to arrive at the result~\footnote{In order to derive the correct result we note that the split in \eqref{eq:M5split} produces the identities, \cite{Witten:1996hc}, $p_1 (TY) = p_1(TZ) + p_1 (TX), p_2(TY) = p_2(TZ) + p_2(TX) +p_1(TZ) p_1(TX)$. In addition, as reviewed in detail in the Appendix, the equivariant pushforward over the Pontryagin classes of the tangent bundle of $Z$ involves additional derivatives of the equivariant volume, on top of the derivatives coming from the Euler-class integration discussed in the previous section.}
\be
\label{eq:I8fromBC}
\begin{split}
\cI_8^{\rm M5}
&:=
\pi_{Z*}\!\left(\cI_{12}^{\rm M}\right)
\\
&=
\frac{N_\text{M5}^3}{24}\,(\omega_1\omega_2)^2
+
\frac{N_\text{M5}}{48}
\left[
\frac14
\bigl(\omega_1^2+\omega_2^2-p_1(TX_8)\bigr)^2
-
(\omega_1\omega_2)^2
-
p_2(TX_8)
\right]\ .
\end{split}
\ee
Under the Chern-Weil map, \eqref{eq:CW1}-\eqref{eq:CW2}, this reproduces the standard M5-brane anomaly polynomial
obtained already in \cite{Harvey:1998bx}.~\footnote{The coefficients above correspond to the classical inflow result and capture the
leading large-$N$ contribution. The exact anomaly coefficients of the interacting
$A_{N_\text{M5}-1}$ $(2,0)$ theory involve the index-theoretic combinations proportional to
$(N_\text{M5}^3-1,N_\text{M5}-1)$
\cite{Intriligator:2000eq,Yi:2001bz,Ohmori:2014kda}.} An analogous $8$-form also underlies the anomaly analysis of six-dimensional $\mathcal N=(1,0)$ SCFTs, including the conformal matter theories engineered from M5-branes probing ADE singularities, which also admit F-theory realizations.~\cite{Ohmori:2014kda,DelZotto:2014hpa} Their anomaly polynomials can be obtained by anomaly matching on the tensor branch and, in appropriate cases, from M-theory anomaly inflow~\cite{Ohmori:2014kda}. It would be interesting to extend the equivariant integration framework to these examples as well.

The advantage of the equivariant formulation is that it allows us to abandon the compact sphere
representative and perform the same computation directly on non-compact toric spaces. In
particular, the natural next step, which was discussed in more detail recently, \cite{Hristov:2026tde}, is to integrate the anomaly polynomial on an arbitrary toric Calabi-Yau four-fold, $X = X_8$:
\be
\label{eq:PT8general}
\cI_0^{\rm M5}
=
(\pi_{X\rightarrow {\rm pt}})_*
\cI_8^{\rm M5}
=
\frac{N_\text{M5}^3}{24}
(\omega_1\omega_2)^2 C_X(\varepsilon)
+
\frac{N_\text{M5}}{24}
\left(
k_1^X k_3^X
-
\omega_1\omega_2\,k_2^X
-
k_4^X
\right)
C_X(\varepsilon)\ ,
\ee
subject to the equivariant Calabi--Yau condition, \eqref{eq:susyintro}
\be
\label{eq:susycond}
k_1^X(\varepsilon)= \sum_i \e_i=\omega_1+\omega_2\ .
\ee
See Appendix \ref{app:equivariant} for the definition of $k_p^X(\varepsilon)$ in terms of equivariant Chern numbers.
This provides the equivariant completion of the M5 anomaly computation. In particular, for
$X=\mathbb C\times \tilde Y$, with $\tilde Y$ a resolved toric Calabi--Yau threefold, it reproduces the
Cardy-limit Sasaki--Einstein index discussed in more detail in
\cite{Hristov:2026tde}. In Appendix \ref{app:A3} we have included some additional examples for resolutions of cones over five-dimensional Sasakian spaces, if the reader is interested in explicitly unpacking the above more condensed formulas.

\subsection{M2-branes and topological strings}
\label{sec:M2}

The M2-brane realization provides the second, and conceptually more recent, interpretation of
the same twelve-dimensional anomaly polynomial. In this case the relevant near-horizon geometry
is described by a Calabi--Yau fourfold cone $N_8=C(SE_7)$,
or, more generally, by a smooth toric Calabi--Yau resolution $X_8$. The four-dimensional factor
entering the anomaly polynomial is now the Euclidean continuation of the AdS$_4$ factor
associated with the M2-brane worldvolume theory. Topologically this space is trivial and we can
represent it equivariantly by
\be
Z_4\simeq \BC^2\ ,
\ee
whose $U(1)^2$ action provides the equivariant parameters entering the computation. Thus, for M2-branes the order is exchanged with respect to the M5 case:
\be
Y_{12}^\text{M2}=Z_4^{(\parallel)}\times X_8^{(\perp)}
\simeq
\BC^2\times X_8\ .
\ee

The M2-brane charge is measured by the flux through the internal space,
\be\label{M2_charge}
\int_{SE_7} *G_4\sim N_\text{M2} ,
\ee
whereas in the grand-canonical ensemble one fixes the chemical potential $\mu_\text{M2}$ conjugate to
$N_\text{M2}$, corresponding to the AdS$_4$ component of the four-form flux,
\be\label{G4_mu}
G_4 \sim \i\, \mu_\text{M2}\,{\rm vol}_{\rm EAdS_4}.
\ee
By the conventions adopted here, the first pushforward of $\cI_{12}$ is therefore performed over the equivariant (Euclidean) AdS$_4$ factor, followed by the integration over the transverse Calabi--Yau fourfold $X$. Thus the two
brane systems are described by very closely related equivariant integration problems,~\footnote{Note that in the absence of magnetic charge, we have no singular behavior of $G_4$, which means the subtleties regarding the excision of a tubular region valid in the M5 description is no longer needed for the stack of M2-branes.} with different choices
of which factor is treated as the first integration space.

The equivariant formulation naturally implements the $\mu$-ensemble by the following ansatz for the four-form~\footnote{The factor of $\i$ is already present in $G_4$ above from the Euclidean continuation of AdS$_4$, and we choose the $2 \pi$ normalization conventionally in order to agree with the one in \cite{Hristov:2026tde}.}
\be
E_4=\frac{\mu_\text{M2}}{2 \pi \i}\,e(T\BC^2)\ .
\ee
The compact Bott--Cattaneo integration appearing in the M5 computation is now replaced by the non-compact equivariant pushforward on a single fixed-point. Explicitly,
\be
\pi_{Z*}
\left((e(T\BC^2))^m\right)
= \partial_{\lam^1} \partial_{\lam^2} \BV_{\BC^2} (\lam, \omega) \Big|_{\lam = 0} =
(\omega_1\omega_2)^{m-1}\ ,
\ee
where $\omega_{1,2}$ are the equivariant parameters of $Z_4\simeq\BC^2$, see \eqref{eq:Cvol}. Applying this
equivariant reduction to the twelve-form \eqref{eq:I12M} gives the intermediate eight-form~\footnote{Apart from the normalization, the remaining calculation coincides exactly with the one leading to \eqref{eq:I8fromBC}, except in this case we only have a single copy of $\BC^2$.}
\begin{align}
\label{eq:M2braneanomaly}
\begin{split}
2\pi \i\, & \cI_8^{\rm M2}
:=
2\pi \i\, \pi_{Z*}(\cI_{12}^{\rm M})
\\
=&
-\frac{\mu_\text{M2}^3}{24 \pi^2}\,(\omega_1\omega_2)^2
+
\frac{\mu_\text{M2}}{48}
\left[
\frac14
\bigl(\omega_1^2+\omega_2^2-p_1(TX_8)\bigr)^2
-
(\omega_1\omega_2)^2
-
p_2(TX_8)
\right]\ ,
\end{split}
\end{align}
where equivariant Calabi--Yau condition again imposes \eqref{eq:susycond}. Note that a particular combination of two equivariant parameters $\omega_{1,2}$ is eventually identified with the thermal circle of Euclidean AdS$_4$ in global coordinates, which imposes a further constraint. Although important for the specific match with the dual $S^3$ partition function and the M2/M5 correspondence, which were more carefully discussed in \cite{Hristov:2026tde}, we will not explicitly need to utilize this here.

The remaining pushforward over $X_8$ is now identical to the one appearing in the M5
calculation:
\be
\label{eq:M2zeroform}
\begin{split}
-2\pi \i\, \cI_0^{\rm M2}(\mu)
&=-2\pi \i\, 
(\pi_{X\rightarrow{\rm pt}})_*\cI_8^{\rm M2}
\\
&=-
\frac{\mu_\text{M2}^3}{24 \pi^2}(\omega_1\omega_2)^2 C_X(\e)
+
\frac{\mu_\text{M2}}{24}
\Big(
k_1^X k_3^X
-\omega_1\omega_2\,k_2^X
-k_4^X
\Big)C_X(\e)\ ,
\end{split}
\ee
which reproduces the finite-$N$ M2-brane free energies obtained in
\cite{Cassia:2025jkr}, and further discussed in \cite{Hristov:2026tde}. Note that the result follows directly from the anomaly polynomial and equivariant geometry. The result \eqref{eq:M2zeroform} also agrees with the supersymmetric M-theory action computed in \cite{BenettiGenolini:2026cyc} in the $\mu$ ensemble directly from 11-dimensional localization, including the overall factor of $C_X(\epsilon)$ for the equivariant volume of $X_8$.

We emphasize that this result is insensitive to metric deformations of the toric manifold under consideration. In particular, it already incorporates the results for the M2 brane free-energies when the near-horizon geometry is (Euclidean) AdS$_4$ with an $S^3$ or squashed $S^3$ boundary, the difference amounting to distinct choices of the equivariant parameters $\omega_1,\omega_2$ for the toric action on the $Z_4$ factor, which can be found for instance in \cite{Cassia:2025jkr}.

\subsection*{Topological strings}
On the other hand, the same expression \eqref{eq:M2zeroform} was previously obtained from a
completely different perspective. In the equivariant topological-string formulation developed in
\cite{Cassia:2025aus,Cassia:2025jkr}, the perturbative grand-canonical M2-brane partition
function is governed by the constant-map sector of the topological string on the toric
Calabi--Yau fourfold $X_8$. The perturbative free energy takes the form
\be
\label{eq:pertstring}
F^{\rm top,pert}_X(\lambda,\varepsilon;g_s)
=
\sum_{\mathfrak g=0}^{\infty}
g_s^{2(\mathfrak g-1)}
\,N^{X}_{\mathfrak g,0}(\lambda,\varepsilon),
\ee
where $\lambda$ denotes the redundant K\"ahler parameters introduced in
\eqref{eq:VXdef}. For the mesonic twist relevant to the M2-brane partition function one
sets\footnote{This is the universal parametrization discussed in \cite{Cassia:2025aus}, which amounts to choosing}
\begin{equation}
\lambda^i=\tilde \mu\ ,
\end{equation}
so that all dependence on the geometry is encoded in the equivariant parameters
$\varepsilon_i$, while the free energy becomes a cubic polynomial in the chemical potential
$\tilde \mu$. More precisely, the genus-zero contribution is cubic in $\tilde \mu$, the genus-one
contribution is linear, and all higher-genus constant maps contribute only $\tilde \mu$-independent
constants. Note that the chemical potential $\tilde \mu$ is conjugate to the exact M2-brane charge, not to the bare number $N_\text{M2}$ that quantizes the flux $*G_4$, \cite{Cassia:2025aus}, which explains the different notation. We come back to this point below.

The remarkable observation of \cite{Cassia:2025jkr} (see also \cite{Hristov:2026zjh}) is that these first two terms already
reproduce the complete perturbative finite-$N$ M2-brane free energy. From the present point of
view this result admits a simple geometric interpretation. Equation~\eqref{eq:M2zeroform}
shows that exactly the same cubic and linear dependence follows directly from the classical
M-theory anomaly polynomial through two successive equivariant pushforwards,
\be
\cI^{\rm M}_{12}
\longrightarrow
\cI^{\rm M2}_{8}
\longrightarrow
\cI^{\rm M2}_{0},
\ee
without any reference to topological strings.

The leading term follows immediately. The cubic contribution is simply the equivariant volume of
the Calabi--Yau fourfold,
\be
(\mu_\text{M2})^3\,(\pi_{X\rightarrow{\rm pt}})_*(1)
=
(\mu_\text{M2})^3\,C_X(\varepsilon),
\ee
which is precisely the genus-zero constant-map contribution of
\cite[Eq.~(2.18)]{Cassia:2025jkr}, upto normalization.
The linear term is equally natural. Using the supersymmetry
condition \eqref{eq:susycond}, the second line of \eqref{eq:M2zeroform} becomes
\be
\frac{\mu_\text{M2}}{24}
\Big(
k_1^Xk_3^X
-\omega_1\omega_2\,k_2^X
-k_4^X
\Big)
C_X(\varepsilon)
=
\frac{\mu_\text{M2}}{24}
\Big[
(\omega_1+\omega_2)c_3^X(\varepsilon)
-\omega_1\omega_2\,c_2^X(\varepsilon)
-c_4^X
\Big],
\ee
where we used $c_p^X=k_p^X\,C_X$. The first two terms reproduce exactly the genus-one
constant-map invariants identified in \cite{Cassia:2025jkr}, while the final contribution,
proportional to the Euler characteristic,
\be
c_4^X=\chi(X)\ ,
\ee
accounts for the familiar quantum shift of the M2-brane charge, \cite{Bergman:2009zh,Cassia:2025aus}
\be
N_\text{M2}\;\rightarrow\;
N_\text{M2}-\frac{\chi(X)}{24}\ ,
\ee
which arises from the precise canonical relation between the K\"ahler parameters and the
chemical potential, $\tilde \mu$. In particular, the number of M2-branes with the above constant shift is the precise charge which is conjugate to $\tilde{\mu}$.

The significance of this agreement extends well beyond a consistency check. The computation
presented here is entirely classical: it starts from the 
anomaly polynomial and uses only equivariant characteristic classes together with equivariant
pushforwards. By contrast, the derivation of \cite{Cassia:2025aus,Cassia:2025jkr} is intrinsically quantum (though constant maps still relate to classical geometry),
being formulated as a worldsheet genus expansion of the topological string. Nevertheless, the
two constructions terminate in exactly the same equivariant expression. 

This viewpoint also places the present results in a broader historical context. Already in the
work of Harvey, Moore and Minasian~\cite{Harvey:1998bx}, the same
12d anomaly was shown, after compactification on a
Calabi--Yau threefold, to determine the central charges governing the entropy of MSW black
strings, ~\cite{Maldacena:1997de}. Through the OSV conjecture~\cite{Ooguri:2004zv}, these same characteristic classes
reappear in the topological-string partition function. The relation uncovered here may therefore
be viewed as the M2-brane analogue of this classical correspondence: once formulated
equivariantly, the anomaly polynomial already contains the complete perturbative information
captured by the constant-map sector of the topological string.

A conceptual picture thus emerges. The anomaly polynomial appears to define a classical
equivariant theory, while the topological-string partition function furnishes its quantum
completion. The constant maps correspond to the semiclassical approximation encoded by the
classical anomaly integral, whereas higher-genus constant maps and worldsheet instantons should
be interpreted as genuine quantum corrections to this underlying equivariant geometry. We hope to explore this connection further in future
work.

\section{Type IIB: D3-brane anomalies}
\label{sec:IIB}

The D3-brane case combines the two viewpoints encountered for M2- and M5-branes into a single
framework. Because the type IIB five-form field strength is self-dual,
\be
F_5=*F_5,
\ee
the anomaly polynomial admits both a magnetic description, in which the flux is measured through
the transverse five-sphere, and an electric description, in which $F_5$ is identified with the
AdS$_5$ volume form. These two descriptions are physically equivalent, but they correspond to the
two opposite choices of equivariant pushforward introduced in the Introduction.

The anomaly-inflow analysis of Bah, Bonetti, Minasian and Weck
\cite{Bah:2020jas} is naturally formulated in terms of the eleven-form~\footnote{Note the different normalization due to the different conventions for the global angular forms used here, c.f.\ subsection \ref{subsec:odd}.}
\be
\cI_{11}=\frac14\, E_5\wedge dE_5,
\ee
where $E_5=N_\text{D3}\,e_5$ is the global angular form of the $SO(6)$ normal bundle, analogous to the M5-brane construction except for the odd dimensionality. Since $E_5$ is not closed, see subsection \ref{subsec:odd}, we write $dE_5=:E_6$ and Stokes' theorem allows one to rewrite the corresponding anomaly integral as a bulk
integral over the 6d disk bounded by the five-sphere,
\be
\int_{S^5}\cI_{11}
=
\int_{D^6} \left(\frac12\, E_6\wedge E_6\right)\ .
\ee

Motivated by this observation, throughout this section we regard
\be
\cI_{12}^{\rm IIB}
=
\frac12\, E_6\wedge E_6
\ee
as the fundamental equivariant object, completely analogous to the M-theory twelve-form
$\cI_{12}^{\rm M}$, except it is strictly applicable to D3-branes and no other type IIB objects. We remind the reader that in what follows we consider the setup
\be
Y_{12}=X_6^{(\parallel)}\times Z_6^{(\perp)}\ .
\ee
Notice that, although we explicitly used $S^5$ and its cone when arguing about the anomaly polynomial, in what follows we are going to relax this condition and allow for both $X$ and $Z$ to be arbitrary toric Calabi-Yau three-folds, or equivalently resolutions of cones over toric Sasakian five-folds.

\subsection{Magnetic D3-branes}
In analogy to the M5-brane case, and in accordance with the discussion in subsection \ref{subsec:odd}, for the magnetic description of the D3-branes one identifies $E_6$ with the equivariant Euler class of the tangent bundle of the transverse space, $Z_6\equiv \mathbb{C}^3$,
\be
\label{eq:D3magansatz}
E_6=N_\text{D3}\,e (T\mathbb \BC^3)\ ,
\ee
where $N_\text{D3}$ is the number of D3-branes.~\footnote{In this section we are only dealing with D3-branes, which means that the brane number $N_\text{D3}$ and the chemical potential $\mu_\text{D3}$ are conjugate to each other, unlike $N_\text{M5}$ and $\mu_\text{M2}$ in the previous section. For this reason, henceforth, we drop the D3 index from both quantities.}
Once again, the tangent bundle may be decomposed as a sum of complex line bundles. In what follows, we will denote by $\omega_i$ the equivariant parameters of the toric action on $Z_6$, and by $\epsilon_i$ the ones for the toric action on $X_6$.

The first equivariant pushforward therefore gives
\be
\cI_6^{\rm mag}
=
\pi_{Z*} \left( \cI_{12}^{\rm IIB} \right)
=
\frac{N^2}{2}
\int_{\mathbb C^3}
\left[e(T\mathbb C^3)^2\right]\ .
\ee
Using the identity \eqref{eq:CEulerclass},
one immediately obtains
\be\label{I_6^mag}
\cI_6^{\rm mag}
=
\frac{N^2}{2}\,
\omega_1\omega_2\omega_3 = \frac{N^2}{2\, C_Z(\omega)}\ .
\ee
For a general toric Calabi--Yau threefold $Z$, which we cannot directly do in this magnetic case since it requires a change in the ansatz \eqref{eq:D3magansatz}, we still expect to recover the latter answer. This corresponds to the familiar inverse-volume relation for the Sasakian volume, underlying
$a$-maximization
\cite{Martelli:2005tp,Martelli:2006yb,Butti:2005vn,Lee:2006ru,Benvenuti:2006xg}, where $C_Z$ above is directly proportional to the Sasakian volume, see \cite{Martelli:2023oqk}. 

One can view this result as a genuine 6-form over the base manifold (more precisely, the evaluation at a fixed point of the equivariant completion thereof) by invoking the Chern-Weil correspondence argued for around eq. \eqref{eq:CW1}. In this way one recovers the relation between the D3-brane anomaly polynomial and $c_3(E)$, the third Chern class of the normal bundle (or R-symmetry bundle), commonly written in the literature, see e.g. \cite{Bah:2020jas}.

Performing the second equivariant pushforward over an auxiliary toric
threefold $X$ therefore gives
\be
\label{eq:D3fullanswerN}
\cI_0^{\rm D3}(N)
=
\frac{N^2}{2}
\frac{C_X(\e)}{C_Z (\omega)}\ ,
\ee
as anticipated in the introduction.

\subsection{Electric D3-branes}

The electric description is completely parallel to the M2-brane computation. Instead of fixing
the magnetic flux through the transverse sphere, one works in the grand-canonical ensemble and from self-duality
identifies~\footnote{The factor of $\i$ comes from the Euclidean continuation, in analogy to the M2-brane discussion.}
\be
E_6
= \i\,
\mu\,e(T \BC^3)\ ,
\ee
where $X_6\simeq\mathbb C^3$ is the six-dimensional extension of Euclidean AdS$_5$.

The first equivariant pushforward is therefore
\be
\label{I_6^el}
\cI_6^{\rm el}
=
\pi_{X*}
\left( \cI_{12}^{\rm IIB} \right)
=
-\frac{\mu^2}{2}
\e_1\e_2\e_3\ = -\frac{\mu^2}{2}\,  \frac{1}{C_X (\e)}\ ,
\ee
and the second pushforward over the internal Calabi--Yau threefold gives
\be\label{eq:D3eCYpf}
\cI_0^{\rm D3}(\mu)
=
- \frac{\mu^2}{2}\, \frac{C_Z (\omega)}{C_X(\e)}\ .
\ee

Passing from the grand-canonical to the canonical ensemble by a Legendre transform yields
\be\label{Laplace_D3}
\cI_0^{\rm D3}(N)
=
\frac{N^2}{2}
\frac{C_X(\e)}{C_Z (\omega)}\ ,
\ee
which reproduces the magnetic answer we already derived, \eqref{eq:D3fullanswerN}.

The D3-brane system therefore provides a particularly transparent realization of the general
picture advocated in this paper. The magnetic and electric descriptions correspond to the two
possible orders of equivariant pushforward of the same twelve-form
$\cI_{12}^{\rm IIB}$, exactly mirroring the M5- and M2-brane constructions discussed in the
previous sections. The difference is that, owing to the self-duality of $F_5$, the two
descriptions are not merely dual but physically equivalent. In this sense, the D3-brane
computation provides the simplest example of the general principle that a single
higher-dimensional equivariant anomaly polynomial can encode distinct lower-dimensional brane
partition functions, depending only on which directions are integrated out first.

\subsection*{Topological strings for D3 brane models}

We also note that the results here also agree with the picture advocated in \cite{Cassia:2025jkr}, based also on \cite{Martelli:2023oqk,Colombo:2023fhu}, for the relation between the partition function of D3 brane models and the topological string expansion; namely if we inverse Laplace transform \eqref{I_6^mag}, we  obtain an equivariant constant map of second degree with the mesonic twist $\lambda^i=\tilde{\mu}$. Explicitly,

\begin{equation}
    \mathcal{I}_0^{\text{D3}}(N)=\int\text{d}\tilde{\mu}\exp\left(2C_Z(\omega)\tilde{\mu}^2-2N\tilde{\mu}\right)\,,
\end{equation}
where $N$ is directly the number of D3-branes, and $F^{\text{top,pert}}_{Z,(2)}(\lambda_{\text{mes.}}=\tilde{\mu},\omega)\equiv 2\, C_Z(\omega)\tilde{\mu}^2$ relates to the second order term in the equivariant topological string expansion \cite{Cassia:2025jkr}. 

This complements the arguments from the previous section that the 12d anomaly polynomial, depending only on quantities that characterise the classical geometry of the brane system, captures the data in topological string constant maps.

\subsection{Comments on general $X^{\parallel}$ and $Z^\perp=C(\text{SE}_5)$}
\label{sec:4.3}

As noted above, the Euler-class ansatz for $E_4$ in the magnetic case for $Z_6^{(\perp)}=\BC^3$  does not extend to arbitrary integration manifolds. Nevertheless, we claim that the form of the anomaly polynomials \eqref{I_6^mag} and \eqref{I_6^el}, as well as the final answers for $\cI_0$, hold for any toric Calabi--Yau. The preceding analysis provides evidence for this expectation. In the electric description, the second pushforward in \eqref{eq:D3eCYpf} is well defined for an arbitrary internal $Z^{(\perp)}_6$. By electric-magnetic self-duality of D3-branes, its Legendre transform must coincide with the magnetic description, in which the first pushforward is over the $Z^{(\perp)}_6$ and the second over an auxiliary toric threefold $X_6^{(\parallel)}$. Thus, although the magnetic ansatz \eqref{eq:D3magansatz} cannot be formulated directly for an arbitrary Calabi--Yau, the self-consistency between the electric and magnetic descriptions bypasses this obstruction.

For readers that find the above discussion somewhat abstract, here we briefly unpack the compact notations used so far. Even though at first sight it might be perplexing, the following general identity holds for the mesonic equivariant volume of an arbitrary CY threefold, $Z$:
\be
\label{eq:cubiccoeff}
    \frac1{C_Z (\omega)} = \sum_{i, j, k} c_{i j k}\, \omega_i\, \omega_j\, \omega_k := \sum_{i, j, k} \Big| \det \begin{pmatrix} v^i \\ v^j \\v^k \end{pmatrix} \Big|\, \omega_i\, \omega_j\, \omega_k\ ,  
\ee
where the latter equality can be seen as the definition of the anomaly (or Chern-Simons) coefficients $c_{i j k}$ that (sometimes upto normalizations) appear both in supergravity and in field theory literature, see \cite{Witten:1998qj,Tachikawa:2005tq,Hristov:2025ygn}. The vectors $v^i$ define the underlying toric fan, the cokernel of the GLSM charges $Q$, see the Appendix for more details. The reason for the above equation is simply the fact that the mesonic equivariant volume matches with the Sasakian volume of the underlying space, \cite{Martelli:2005tp,Martelli:2006yb,Martelli:2023oqk}. The resulting cubic anomaly coefficients precisely agree with those of the D3-brane worldvolume theory derived in \cite{Benvenuti:2006xg} for any toric SE$_5$ manifold, as proven in \cite{Butti:2005vn,Lee:2006ru}. In addition, we have presented several common examples of Sasakian cones in Appendix\ \ref{app:A3}, including the resolved conifold, $\cC$, and resolutions of the cones over $Y^{p,q}$ and $L^{p,q,r}$ manifolds, where one can find more explicit details about how to read off the cubic coefficients $c_{i j k}$ from the toric geometry.

\subsection*{On choices of $X^{(\parallel)}$}

We finish with a comment on the choice of worldvolume directions $X^{(\parallel)}$. Since we have so far only looked at direct product of the two manifolds, $X$ and $Z$, we can look at cases where the spacetime background does not feature magnetic charges. A prominent class of solutions of this kind are the Gutowski-Reall black holes and generalizations, \cite{Gutowski:2004yv,Chong:2005hr,Kunduri:2006ek}, whose holographic dual is simply the superconformal index of the D3-brane theory, \cite{Kinney:2005ej,Romelsberger:2005eg}. The construction of the extension space $X$ in this case is straightforward, since we need to extend the product of a circle $S^1$ with a three-sphere, $S^3$. The natural choice in out setting is to simply take the product of the two cones, $X^{(\parallel)} = \BC \times \BC^2$, \cite{Cassani:2024tvk,Hristov:2026tde}. We should further keep notice of the equivariant parameter associated with the thermal circle (we can call it $\e_3$, leaving $\e_{1,2}$ for the $S^3$ directions), which by supersymmetry is fixed to a constant, $\e_3 = \pm 2 \pi \i$, see \cite{Cassani:2024tvk}. In the $N$-ensemble, for example, we then simply arrive at the following general formula,
\be
\cI_0^{\rm D3}(N)
=
\frac{N^2}{2}
\frac{C_X(\e)}{C_Z (\omega)} = \frac{N^2}{2}
\frac{\sum_{i, j, k} c_{i j k}\, \omega_i\, \omega_j\, \omega_k}{2 \pi \i\, \e_1 \e_2}\ .
\ee
This expression (up to varying conventions) correctly reproduces the known grand-canonical entropy function of the supersymmetric black holes in AdS$_5$, and the dual large-$N$ superconformal indices, \cite{Hosseini:2017mds,Cabo-Bizet:2018ehj,Choi:2018hmj,Benini:2018ywd}. The generalization to $X = \BC \times \BC^2/\mathbb{Z}_p$, which corresponds to Lens space indices, is straightforward.

We expect that a generalization of our ansatz can accommodate backgrounds with non-trivial magnetic fluxes through non-trivial cycles. We plan to return to this point, together with the other interesting directions mentioned below.

\section{Discussion and outlook}
\label{sec:discussion}

We have shown that the anomaly-polynomial integrals underlying the partition functions of
M2-, M5-, and D3-brane systems admit a unified reformulation in terms of equivariant volumes and
equivariant Euler-class integrals. From this viewpoint, the classical Bott--Cattaneo formula for
sphere bundles appears as the compact prototype of a more general equivariant construction:
the fiber integral is reproduced by gluing the local contributions from the fixed points, with
opposite equivariant parameters. This perspective explains the relation between the M5-brane
anomaly-inflow computation and the M2-brane equivariant topological-string description found in
\cite{Hristov:2026tde}, and extends naturally to D3-branes, where the same equivariant structure
underlies the relation between $a$-maximization and volume minimization.

Several aspects of this framework deserve further investigation.

\medskip

\noindent{\bf Equivariant localization and supergravity.}
The recent progress on equivariant localization of supersymmetric supergravity actions suggests
that the equivariant integrals studied here should admit a direct interpretation from the
corresponding eleven-dimensional and type IIB supergravity path integrals. At present, however,
the precise relation between these two approaches remains to be established. In particular, the
recent localization results of
\cite{Couzens:2026xmi,BenettiGenolini:2026cdw,BenettiGenolini:2026cyc}
involve a careful treatment of boundary contributions but some conjectural simplifications involving holographic renormalization remain unproven. Conversely,
the equivariant-volume approach provides a natural mathematical framework for defining
integrals over the non-compact spaces that arise in these problems. Establishing the precise
equivalence between supergravity localization \emph{including} holographic renormalization and equivariant anomaly integrals should provide a
first-principles derivation of the formulas appearing throughout this work.

\medskip

\noindent{\bf From classical anomalies to quantum partition functions.}
A particularly intriguing aspect of the present construction is that the starting point is entirely
classical: the twelve-dimensional anomaly polynomial is a characteristic class determined by
the classical supergravity fields and the brane charges (even though it does incorporate higher-derivative corrections in supergravity). Nevertheless, after equivariant
integration it reproduces quantities that are usually regarded as genuinely quantum, such as
the perturbative part of supersymmetric partition functions and, on the M2-brane side, the
constant-map sector of equivariant topological strings.

This suggests that anomaly inflow may provide more than a computational shortcut for known
protected quantities. Rather, it may identify the geometric origin of the universal quantum
contributions that survive in the large-$N$ limit. In particular, the agreement between the
M-theory anomaly integral and the constant-map contribution of the topological string indicates
that the equivariant anomaly polynomial captures the part of the quantum partition function
which is insensitive to the detailed enumerative geometry, while the remaining instanton
sectors should correspond to further geometric data beyond the characteristic classes appearing
in the present construction.

Understanding whether the full quantum partition functions can be organized as an equivariant
completion of the anomaly polynomial, with constant maps providing the leading universal piece,
is an interesting open problem. Such a perspective could provide a bridge between anomaly
inflow, equivariant localization, and the enumerative structures of topological string theory.

\medskip

\noindent{\bf Beyond factorized flat brane geometries.}
The anomaly polynomials considered in this paper,
\eqref{eq:Manom} and \eqref{eq:Banom},
are the natural inflow polynomials associated with M-theory and type IIB brane systems in the
setups of interest here. Our analysis has focused on configurations where the relevant geometry
admits a simple factorization into transverse and worldvolume equivariant directions. More
general brane configurations, such as wrapped branes, defects, and intersecting brane systems,
involve additional geometric data associated with the embedding, twisting, and localized degrees
of freedom. In the M5-brane case, systematic anomaly inflow constructions for wrapped
configurations have been developed in
\cite{Bah:2018jrv,Bah:2019jts,Bah:2019rgq},
and analogous structures appear for D3-brane systems in
\cite{Bah:2020jas}.
We expect that all M-brane systems are still described by the same 12d anomaly, but different $E_4$ ans\"atze and underlying manifolds, while more general IIB systems will also require a more general set of 12 anomaly forms, see \cite{Couzens:2026xmi}. It is interesting to repeat here our observation that our electric brane calculations (ignoring the subleading piece in the M-theroy description) eventually could be matched with the extremization procedure in \cite{Martelli:2023oqk,Colombo:2023fhu}. We expect that this continues to be the case for wrapped brane configurations, discussed there.

\medskip

\noindent{\bf Exact duality pattern?}
The structural similarity between the M2/M5 relation discussed in Section~\ref{sec:Mtheory}
and the electric/magnetic D3-brane relation of Section~\ref{sec:IIB} suggests that these
examples may be manifestations of a broader organizing principle. In both cases, apparently
different brane partition functions arise from different interpretations of the same
higher-dimensional equivariant integral: directions that play the role of transverse geometry in
one description become part of the effective worldvolume geometry in another.

This structure has an AGT-like flavor, in the sense that a single higher-dimensional object
admits different lower-dimensional realizations depending on the order in which equivariant
pushforwards are performed. Exploring whether similar structures persist for more general (including
non-conformal) brane systems in type IIB and beyond, is a natural direction for future work. What is perhaps most important is the investigation of whether further perturbative and non-perturbative corrections (genuinely quantum from the geometric point of view) also obey the suggested dualities here, or the present observations hold simply at the level of classical geometry (which still gives finite $N$ results, as discussed).

\acknowledgments
We would like to thank Yi Pang and Gabriele Tartaglino-Mazzucchelli for useful comments on the draft. K.H.\ is supported in part by the Bulgarian NSF grant KP-06-N88/1. P.V.M.\ is supported by the Netherlands Organisation for Scientific Research (NWO) under the VICI grant VI.C.202.104. K.H.\ would like to thank the Raybould fellowship and the School of Mathematics and Physics in UQ for the financial support and hospitality, as well as the organizers of the GGI program "Pathways to Quantum Black Holes" in Florence and the "Eurostrings 2026" conference in Athens for the financial support and stimulating environment.

\appendix

\section{Equivariant geometry}
\label{app:equivariant}

This appendix summarizes the equivariant-geometric conventions used throughout the paper,
following
\cite{Cassia:2025aus,Martelli:2023oqk,Hristov:2026tde},
and collects a number of identities used repeatedly in the main text.

\subsection{Equivariant volumes and characteristic classes}

Let
\[
X=\BC^n/\!\!/U(1)^r
\]
be a smooth toric Calabi--Yau manifold of complex dimension
$d=n-r$. The action of the $U(1)^r$ torus is described in terms of a matrix of integer charges $Q_i^a$ with $i=1,\dots,n$ and $a=1,\dots,r$, corresponding to an embedding of $U(1)^r$ into the larger torus $U(1)^n$.
Its equivariant volume is defined by
\be
\label{eq:AV}
\BV_X(\lambda,\varepsilon)
=
\int_X e^{\omega^{\BT}_\lambda}
=
\oint_{\rm JK}
\prod_{a=1}^{r}
\frac{d\phi_a}{2\pi i}
\,
\frac{e^{x_i\lambda^i}}
{\prod_{i=1}^{n}x_i},
\qquad
x_i=\varepsilon_i+\phi_aQ_i^a,
\ee
where the contour is specified by the Jeffrey--Kirwan prescription. In addition, we can define a map $v:\BR^n\to\BR^{n-r}$ as the cokernel of $Q$, i.e.\
\be
\label{eq:map-v}
 v^i_\alpha Q_i^a = 0\,.
\ee
It can be proven, c.f.\ \cite{Cassia:2025aus}, that the function $\BV_X(\lam,\e)$ depends only on the combinations
\be
\label{eq:nuinsteadofeps}
 \nu_\alpha(\e) := v^i_\alpha\e_i
\ee
which can be regarded as equivariant parameters for the quotient torus $U(1)^n/U(1)^r\cong U(1)^{n-r}$. We give some non-trivial examples of this construction in App.\ \ref{app:A3}.

The equivariant intersection numbers are obtained by differentiation with respect to the
(redundant) K\"ahler parameters,
\be
C^X_{i_1\cdots i_p}(\varepsilon)
:=
\left.
\frac{\partial^p\BV_X}
{\partial\lambda^{i_1}\cdots\partial\lambda^{i_p}}
\right|_{\lambda=0},
\ee
while the mesonic equivariant volume is simply
\be
C_X(\varepsilon)
:= (\pi_{X \to \text{pt}})_* (1) =
\BV_X(0,\varepsilon).
\ee

The
$x_i$ appearing above
denote the equivariant Chern roots of a complex vector bundle $E$.
The total Chern and Pontryagin classes are
\be
c(E)
=
\prod_i(1+x_i)\ ,
\qquad
p(E)
=
\prod_i(1+x_i^2)\ .
\ee

Consequently,
\be
c_1=\sum_i x_i\ ,\qquad
c_2=\sum_{i<j}x_ix_j\ , \qquad
c_3=\sum_{i<j<k}x_ix_jx_k,
\ee
and
\be
p_1=c_1^2-2c_2\ , \qquad
p_2=c_2^2-2c_1c_3+2c_4\ .
\ee

Explicitly, the equivariant Chern numbers are~\footnote{We drop the superscript $\BT$ in what follows.}
\be
c_p^X(\varepsilon)
=
\sum_{i_1<\cdots<i_p}
C^X_{i_1\cdots i_p}(\varepsilon),
\ee
and it is convenient to normalize them by the equivariant volume,
\be\label{eq:kpX}
k_p^X(\varepsilon)
:=
\frac{c_p^X(\varepsilon)}
{C_X(\varepsilon)}.
\ee

For a toric Calabi--Yau manifold (without imposing yet an \emph{equivariant} condition) one has
\be
k_1^X(\varepsilon)
=
\sum_i\varepsilon_i\ ,
\ee
which is simply the equivariant form of the Calabi--Yau condition
$c_1(TX)=0$.

The equivariant Pontryagin classes are expressed in terms of the normalized equivariant Chern numbers \eqref{eq:kpX} as
\begin{equation}\label{eq:p12}
  \begin{aligned}  p_1^X(\varepsilon)&=C_X(\varepsilon)\left((k_1^X(\varepsilon))^2-2k_2^X(\varepsilon)\right)\,,\\
  p_2^X(\varepsilon)&=C_X(\varepsilon)\left((k_2^X(\varepsilon)^2-2k_1^X(\varepsilon)k_3^X(\varepsilon)+2k_4^X(\varepsilon)\right)\,.
\end{aligned}
\end{equation}

For
$\BC^2$, where $Q=0$,
one straightforwardly finds
\bea
c_1(T\BC^2)
&=&
\frac{\e_1+\e_2}{\e_1 \e_2}\ , \qquad
c_2(T\BC^2)
=
1\ ,
\\
p_1(T\BC^2)
&=&
\frac{\e_1^2+\e_2^2}{\e_1 \e_2}\ , \qquad
p_2(T\BC^2)
=
\e_1\e_2\ .
\eea

Similarly, for
$\BC^3$,
\bea
c_1(T\BC^3) &=& \frac{\e_1+\e_2+\e_3}{\e_1 \e_2 \e_3}\,, \quad c_2(T\BC^3) = \frac{\e_1\e_2 +\e_1\e_3 +\e_2\e_3}{\e_1 \e_2 \e_3}\,, \quad c_3(T\BC^3) = 1\,, \\
p_1(T\BC^3) &=& \frac{\e_1^2+\e_2^2+\e_3^2}{\e_1 \e_2 \e_3}\ , \qquad p_2(T\BC^3) = \frac{(\e_1\e_2)^2 +(\e_1\e_3)^2 +(\e_2\e_3)^2}{\e_1 \e_2 \e_3}\ .
\eea
One can reconstruct analogous expressions for any $\BC^n$ simply using the general form of the equivariant volume, c.f.\ \eqref{eq:Cvol}.

\subsection{Some non-trivial examples of local threefolds}
\label{app:A3}

There is a potentially confusing point that cannot be simply illustrated on the $\BC^m$ examples due to their topological triviality. The equivariant parameters $\e$ are in general redundant and overparametrize the equivariant volume, see the statement above \eqref{eq:nuinsteadofeps}. One then finds meaningful additional constraints among the parameters, dubbed mesonic constraints in \cite{Cassia:2025aus}, see \cite{Hosseini:2019use,Hosseini:2019ddy}, that simply amount to imposing~\footnote{Note that we can relax the condition that the right hand side below is evaluated at vanishing $\lam$-parameters, but this does not amount to a change in the equation itself.}
\be
\label{eq:mesonicconstraint}
 \e_i \stackrel{!}{=}
 \left.\frac{\partial}{\partial\lam^i}\log\BV(\lam,\e)\right|_{\lam=0}\ .
\ee
Alternatively, we can skip this step and directly change basis to a set of faithful equivariant parameters, $\nu_i$, as in \eqref{eq:nuinsteadofeps}. Below we give explicitly the example of the conifold $\cC$, as well as the cones over the Sasakian spaces $Y^{p,q}$ and $L^{p,q,r}$, which often appear in holographic applications.

\subsubsection*{Resolved conifold, $C (T^{1,1})$}
A prototipical example is the resolved conifold, $\cC$, which is a threefold corresponding to the resolution of the cone over $T^{1,1}$, obtained as a symplectic quotient of $\BC^4$. The theory of multiple D3-branes on the conifold is known as Klebanov-Witten theory, \cite{Klebanov:1998hh}. In this construction, c.f.\ \eqref{eq:VXdef}, the GLSM charges are given by
\be\label{eq:GLSMT11}
    Q_{\cC} = (1 \quad 1\quad -1 \quad -1)\ .
\ee
It admits two related resolutions, see \cite{Cassia:2025aus} and further details in \cite{Hristov:2026zjh}, and for concreteness we pick one of them:
\be\label{V-}
 \BV_\cC (\lam, \e) = \frac1{(\e_2-\e_1)}\, \left(
 \frac{\mathe^{\lam^2 (\e_2-\e_1)+\lam^3 (\e_3+\e_1) +\lam^4 (\e_4+\e_1)}}
 {(\e_3+\e_1) (\e_4+\e_1)}
 - \frac{\mathe^{\lam^1 (\e_1-\e_2)+\lam^3 (\e_3+\e_2) +\lam^4 (\e_4+\e_2)}}
 {(\e_3+\e_2) (\e_4+\e_2)} \right)\ .
\ee
It is then straightforward to find the mesonic volume, \eqref{eq:mesonicvol},
\be
    C_{\cC} (\e) = \frac{\sum_{i=1}^4\e_i}
 {(\e_1+\e_3) (\e_2+\e_3) (\e_1+\e_4) (\e_2+\e_4)}\ .
\ee
However, at this stage we should observe that the equivariant parameter $\e$ are not associated with the remaining toric action on the conifold itself, but with the original $\BC^4$ geometry. In order to compare with both supergravity and field theory results, we need to impose the constraints \eqref{eq:mesonicconstraint}, which in this case can be shown to simplify to a single additional relation, \cite{Cassia:2025aus},
\be
    \e_1 \e_2 = \e_3 \e_4\ .
\ee
Remarkably, imposing this additional constraint leads to the following identity,\footnote{This can be seen, for instance, by setting $(\epsilon_1\epsilon_2)^2=(\epsilon_3\epsilon_4)^2=\prod_i\epsilon_i$ when writing out the denominator of $C_\mathcal{C}(\epsilon)$; this identity is then explicit.}
\be
    \frac1{C_\cC (\e)} = \e_1 \e_2 \e_3 + \e_1 \e_2 \e_4+\e_1 \e_3 \e_4 + \e_2 \e_3 \e_4 = \frac16\, \sum_{i, j, k} \Big| \det \begin{pmatrix} v^i \\ v^j \\v^k \end{pmatrix} \Big|\, \e_i \e_j \e_k\ ,
\ee
such that the cubic coefficients defined in \eqref{eq:cubiccoeff} are given by
\be
    c_{123} = c_{124} = c_{134} = c_{234} = 1\ ,
\ee
and permutations; where we made use of the toric fan vectors, \cite{Cassia:2025aus}, 
\be
 v_\cC =
 \begin{pmatrix}
 1 & 1 & 0 \\
 1 & 0 & 1 \\
 1 & 1 & 1 \\
 1 & 0 & 0
 \end{pmatrix}\ .
\ee

Alternatively, we can pass to faithful variables $\nu_i$ via the change of variables \eqref{eq:nuinsteadofeps},
\be
\label{eq:conifoldidentity}
    \nu_1 = \e_1+\e_2+\e_3+\e_4\ ,
 \quad
 \nu_2 = \e_1+\e_3\ ,
 \quad
 \nu_3 = \e_2+\e_3\ .
\ee
The mesonic equivariant volume then simply becomes
\be
    C_\cC (\nu) = \frac{\nu_1}{\nu_2 \nu_3 (\nu_1 - \nu_2) (\nu_1 - \nu_3)} \ ,
\ee
where the dependence on the fourth variable in the $\e$-parametrization automatically disappears, without the need of imposing \eqref{eq:mesonicconstraint}. 

\subsubsection*{Resolved $C(Y^{p,q})$}
More briefly, we can extend the conifold example to the general class of cones over $Y^{p,q}$ spaces, defined by the GLSM charge vector, \cite{Gauntlett:2004hh,Martelli:2004wu,Benvenuti:2004dy}
\be
    Q = (p \quad p\quad -p+q \quad -p-q)\ ,
\ee
where $p > q \geq 0$ ($q=0$ is the limit to the conifold example), ${\rm gcd} (p,q) = 1$.
We then find the equivariant volume is given by
\bea
\begin{split}
 \BV (\lam, \e) = \frac{p}{(\e_2-\e_1)}\, \Big(&
 \frac{\mathe^{\lam^2 (\e_2-\e_1)+\lam^3 (p \e_3+(p-q) \e_1)/p +\lam^4 (p\e_4+ (p+q)\e_1)/p}}
 {(p \e_3+ (p-q) \e_1) (p\e_4+(p+q) \e_1)} \\
 &- \frac{\mathe^{\lam^1 (\e_1-\e_2)+\lam^3 (p \e_3+(p-q)\e_2)/p +\lam^4 (p\e_4+(p+q)\e_2)/p}}
 {(p\e_3+(p-q)\e_2) (p\e_4+(p+q)\e_2)} \Big)\ ,
 \end{split}
\eea
which reproduces the answer for the conifold in the case $(p=1, q=0)$. For the cone over any $Y^{p,0}$, the equivariant volume takes the same form as in \eqref{V-} (in the chosen resolution), since the corresponding GLSM charge is just a rescaling by $p$ of \eqref{eq:GLSMT11}, leaving the defining condition of the toric fan vectors \eqref{eq:map-v} unchanged. For general $p$ and $q$ the toric fan is given by
\be
 v =
 \begin{pmatrix}
 1 & (p-q) & 0 \\
 1 & 0 & (p-q) \\
 1 & p & p \\
 1 & 0 & 0
 \end{pmatrix}\ ,
\ee
such that
\be
    c_{123} = p^2-q^2\ , \quad  c_{124} = (p-q)^2\ , \quad c_{134} = p (p-q)\ , \quad c_{234} = p (p-q)\ ,
\ee
and permutations.

\subsubsection*{Resolved $C(L^{p,q,r})$}
A further generalization is picking the GLSM charge vector, \cite{Cvetic:2005ft,Butti:2005sw}
\be
    Q = (p \quad q\quad -r \quad r-p-q = -s)\ ,
\ee
where $q \geq p > 0$, $p+q > r > 0$, and $(p, q)$ coprime with $(r, s)$ above are coprime
leading to the equivariant volume
\bea
\begin{split}
 \BV (\lam, \e) = \frac{1}{(p \e_2-q \e_1)}\, \Big(&
 \frac{p^2\, \mathe^{\lam^2 (p \e_2-q \e_1)/p+\lam^3 (p \e_3+r \e_1)/p +\lam^4 (p\e_4+ (p+q-r)\e_1)/p}}
 {(p \e_3 + r \e_1) (p\e_4+(p+q-r) \e_1)} \\
 &- \frac{q^2\, \mathe^{\lam^1 (q\e_1-p\e_2)/q+\lam^3 (q \e_3+r \e_2)/q +\lam^4 (q\e_4+(p+q-r)\e_2)/q}}
 {(q\e_3+r \e_2) (p\e_4+(p+q-r)\e_2)} \Big)\ .
 \end{split}
\eea
This reporduces the answer for $Y^{p,q}$ when taking the appropriate restriction of parameters. The toric fan in this case is given by
\be
 v =
 \begin{pmatrix}
 1 & r & 0 \\
 1 & 0 & r \\
 1 & p & q \\
 1 & 0 & 0
 \end{pmatrix}\ ,
\ee
such that
\be
    c_{123} = r (p+q-r)\ , \quad  c_{124} = r^2\ , \quad c_{134} = q r\ , \quad c_{234} = p r\ ,
\ee
and permutations.

\bibliographystyle{JHEP}
\bibliography{refs.bib}

\end{document}